\documentclass[a4paper,11pt]{article}

\usepackage{contribution}

\newcommand{\weblink}[2][]{%
    \ifthenelse{\equal{#1}{}}%
    {\textnormal{\url{#2}}}%
    {\textnormal{\href{#2}{#1}}}%
}

\newcommand{\acknowledgements}[1]{%
  \bigskip\bigskip
  \textsf{\textbf{\Large Acknowledgements}} \\[2ex]
  {#1}
  \bigskip
}

\def\beq{\begin{equation}}
\def\eeq#1{\label{#1}\end{equation}}
\def\eeqn{\end{equation}}

\def\beqa{\begin{eqnarray}}
\def\eeqa#1{\label{#1}\end{eqnarray}}
\def\eeqan{\end{eqnarray}}

\let\bar=\overbar

\def\Dslash{\not{\hbox{\kern-4pt $D$}}}
\def\dslash{\not{\hbox{\kern-2pt $\del$}}}

\def\msb{{\bar{\ssstyle M \kern -1pt S}}}

\newcommand{\contribution}[7][]{%
  \clearpage
  \thispagestyle{plain}
  \ifthenelse{\equal{#1}{}}
  {\hypersetup{pdftitle={#2}}}
  {\hypersetup{pdftitle={#1}}}
  \hypersetup{pdfauthor={{#3} {#4}}}
  {\centering\normalfont\LARGE\bfseries\sffamily #2 \par\nobreak}
  \lhead{}
  \chead{%
    \textit{\footnotesize XIV International Conference on Hadron Spectroscopy
      (\weblink[\textit{hadron2011}]{http://www.hadron2011.de}), 13-17 June 2011, Munich, Germany}%
  }
  \rhead{}
  \bigskip
  \begin{center}
    {#3} {#4}\ifthenelse{\equal{#6}{}}{}{\footnote{\weblink[#6]{mailto:#6}}}
    \ifthenelse{\equal{#7}{}}{}{#7} \\
    \textit{#5}
  \end{center}
  \bigskip
}

\renewcommand{\abstract}[1]{%
  \begin{center}
    \begin{minipage}{0.85\textwidth}
      \begin{footnotesize}
        #1
      \end{footnotesize}
    \end{minipage}
  \end{center}
  \bigskip
}

\makeatletter
\@ifundefined{c@affiliation}%
{\newcounter{affiliation}}{}%
\makeatother
\newcommand{\affiliation}[2][]{\setcounter{affiliation}{#2}%
  \ensuremath{{^{\alph{affiliation}}}\text{#1}}}
\def\tstrutb{\vrule height2.5ex depth0pt width0pt} 

\begin{document}

%
%
%
%
%
{  


%

\contribution[Microscopic Model of Charmonium Strong Decays]
{Microscopic Model of Charmonium Strong Decays}
{J.}{Segovia}  
{\affiliation[Departamento de F\'isica Fundamental and IUFFyM]{1} \\
 Universidad de Salamanca, E-37008 Salamanca, Spain}
{segonza@usal.es}
{\!\!$^,\affiliation{1}$, 
D.R. Entem\affiliation{1}, and 
F. Fern\'andez\affiliation{1}}
%

\abstract{%
Although the spectra of heavy quarkonium systems has been successfully explained
by certain QCD motivated potential models, their strong decays are difficult to
deal with. We perform a microscopic calculation of charmonium strong decays
using the same constituent quark model which successfully describes the
$c\bar{c}$ meson spectrum. We compare the numerical results with the $^{3}P_{0}$
and the experimental data. Comparison with other predictions from similar models
are included.
}
%

\section{Introduction}

Meson strong decay is a complex non-perturbative process that has not yet been
described from QCD first principles. Instead, several phenomenological
models have been developed to deal with this topic, like the $^{3}P_{0}$
model~\cite{mic69_1}, the flux-tube model~\cite{kok87_1}, or microscopic
models (see Refs.~\cite{Eichten78,Eichten06,Swanson96}). The difference between
the two approaches lies on the description of the $q\bar{q}$ creation vertex.
While the $^{3}P_{0}$ model assumes that the $q\bar{q}$ pair is created from the
vacuum, in the microscopic models the $q\bar{q}$ pair is created from the
interquark interactions acting in the model.

The main ingredients in both calculations are the one-gluon exchange and the
linear confinement. The differences lie in the Lorentz structure of the
confinement being vector for Ref.~\cite{Eichten78,Eichten06} and scalar for
Ref.~\cite{Swanson96}. Phenomenology suggests that confinement has to be
dominantly scalar in order to reproduce the hyperfine splitting observed in the
charmonium sector. Strong decays may provide a new physics of information about
the Lorentz structure.

In the present work, we generalize the schematic microscopic models of
Refs.~\cite{Eichten78,Eichten06,Swanson96} using a more realistic constituent
quark model which includes a linear screened confinement and studying the
possible influence of the mixture of scalar and vector Lorentz structures.

\section{Constituent quark model}

Constituent quark masses, coming from the spontaneous chiral symmetry breaking
of the QCD Lagrangian, together with the perturbative one-gluon exchange (OGE)
and the non-perturbative confining interaction are the main pieces of potential
models. In a pure gluon gauge theory the potential energy of the $q\bar{q}$ pair
grows linearly with the interquark distance. However, the presence of sea quarks
may soften the linear potential. Using this idea, Vijande {\it et
al.}~\cite{Vijande2005} developed a model which is able to describe meson
phenomenology from the light to the heavy quark sector. This model incorporates
a confinement potential $V_{\rm CON}^{\rm scalar}(\vec{r}_{ij}) = V_{\rm
CON}^{\rm vector}(\vec{r}_{ij})=\left[-a_{c}(1-e^{-\mu_{c}r_{ij}})+\Delta
\right] (\vec{\lambda}_{i}^{c}\cdot\vec{\lambda}_{j}^{c})$, with a mixture
of a scalar and vector Lorentz structures $V_{\rm CON}(\vec{r}_{ij})=a_{s}V_{\rm
CON}^{\rm scalar}(\vec{r}_{ij})+(1-a_{s})V_{\rm CON}^{\rm
vector}(\vec{r}_{ij})$.

To evaluate the strong decay amplitudes, we solve the Schr\"odinger equation
using the Gaussian Expansion Method~\cite{Hiyama2003}. The model parameters can
be found in Ref.~\cite{Segovia2008}.

\section{A microscopic decay model}

In the microscopic decay models the interaction Hamiltonian can be written
as~\cite{Swanson96}
\begin{equation}
H_{I}=\frac{1}{2} \int d^{3}\!xd^{3}\!y\,J^{a}(\vec{x})K(|\vec{x}-\vec{y}|)J^{a}
(\vec{y}).
\label{Hint}
\end{equation}
The current $J^{a}$ in Eq.~(\ref{Hint}) is assumed to be a color octet. The
currents $J$ (with the color dependence $\lambda^{a}/2$ factored out) are
$J(\vec{x})=\bar{\psi}(\vec{x})\Gamma\psi(\vec{x})$ where $\Gamma={\cal
I},\,\gamma^{0},\,\vec{\gamma}$. The kernels associated with the currents
described before are
$K(r)=-4a_{s}\left[-a_{c}(1-e^{-\mu_{c}r})+\Delta\right],\,+\frac{\alpha_{s}}
{r} \mbox{ and }-\frac{\alpha_{s}}{r}$. For the vector Lorentz structure of the
confinement we use as a kernel
$K(r)=\pm(1-a_s)4\left[-a_{c}(1-e^{-\mu_{c}r})+\Delta\right]$, where $\pm$
refers to static and transverse vector terms, respectively.

\section{Results and conclusions}

The predictions for the total decay rates using the $^{3}P_{0}$ and the
microscopic model are shown in Table~\ref{tab:Tdecayrates}. In general the
total widths are lower in the microscopic model without improving the agreement
with the experimental data.

\begin{table}[t!]
\begin{center}
\begin{tabular}{ccrcc}
\hline
State & $^{3}P_{0}$ & Mic. & Ref.~\cite{PDG2010} & Ref.~\cite{Set} \\
\hline
$\psi(3770)$ & $26.4$ & $19.0$ & $27.6\pm1.0$ & \\
$\psi(4040)$ & $111.0$ & $39.1$ & $80\pm10$ & \\
$\psi(4160)$ & $115.7$ & $32.7$ & $103\pm8$ & \\
$X(4360)$ & $113.7$ & $102.2$ & $74\pm15\pm10$ & \\
$\psi(4415)$ & $115.7$ & $42.7$ & $62\pm20$ & $119\pm16$ \\
$X(4630)$ & $206.0$ & $188.2$ & $92^{+40+10}_{-24-21}$ & \\
$X(4660)$ & $134.8$ & $142.2$ & $48\pm15\pm3$ & \\
\hline
\end{tabular}
\caption{\label{tab:Tdecayrates} Total decay rates, in MeV, predicted by the
$^{3}P_{0}$ and the microscopic models.}
\end{center}
\end{table}

It is difficult to compare our results with former calculations because either
they are not fitted to the heavy quark sector~\cite{Swanson96} or does not
include the same pieces of the current~\cite{Eichten78,Eichten06}. For the sake
of the comparison we show in Table~\ref{tab:compj0Kj0} the results of
Ref.~\cite{Eichten06} together with our model prediction including only the
static vector contribution and the full decay model. The basic difference
between the two calculations is that in Ref.~\cite{Eichten06} the coupling with
the meson-meson channels is treated nonperturbatively and this enhances the
results when the threshold is close to the state. The predictions of the full
decay model are clearly below the experimental data.

\begin{table}[t!]
\begin{center}
\begin{tabular}{lcccc}
\hline
Decay & Ref.~\cite{Eichten06} & $j^{0}Kj^{0}$ & Mic. & Exp.~\cite{PDG2010} \\
\hline
$\psi(3770)\to DD$ & $20.1$ & $29.8$ & $19.0$ & $27.6\pm1$ \\
\hline
$\psi(4040)\to DD$ & $0.1$ & $1.4$ & $10.2$ & \\
$\psi(4040)\to DD^{\ast}$ & $33.0$ & $25.2$ & $18.7$ & \\
$\psi(4040)\to D^{\ast}D^{\ast}$ & $33.0$ & $35.0$ & $9.1$ & \\
$\psi(4040)\to D_{s}D_{s}$ & $8.0$ & $0.3$ & $1.1$ & \\
total & $74.0$ & $61.9$ & $39.1$ & $80\pm10$ \\
\hline
$\psi(4160)\to DD$ & $3.2$ & $25.0$ & $17.0$ & \\
$\psi(4160)\to DD^{\ast}$ & $6.9$ & $0.5$ & $7.4$ & \\
$\psi(4160)\to D^{\ast}D^{\ast}$ & $41.9$ & $21.3$ & $5.3$ & \\
$\psi(4160)\to D_{s}D_{s}$ & $5.6$ & $0.03$ & $2.6$ & \\
$\psi(4160)\to D_{s}D_{s}^{\ast}$ & $11.0$ & $0.6$ & $0.4$ & \\
total & $69.2$ & $47.4$ & $32.7$ & $103\pm8$ \\
\hline
\end{tabular}
\caption{\label{tab:compj0Kj0} Decay rates, in MeV, reported in
Ref.~\cite{Eichten06} and our decay rates taking into account only the static
vector contribution and the full model.}
\end{center}
\end{table}

Finally, in Table~\ref{tab:ratios} we compare the experimental ratios of some
charmonium decays with the prediction of the different models. None of them can
explain the experimental data.

\begin{table}[t!]
\begin{center}
\begin{tabular}{cccccccc}
\hline
\tstrutb
State & Ratio & Cornell & $j^{0}Kj^{0}$ & Mic. & $^{3}P_{0}$ &
Measured~\cite{PDG2010} \\
\hline
\tstrutb
$\psi(4040)$ & $D\bar{D}/D\bar{D}^{\ast}$ & $0.003$ & $0.06$ & $0.54$ &
$0.21$ & $0.24\pm0.05\pm0.12$ \\
& $D^{\ast}\bar{D}^{\ast}/D\bar{D}^{\ast}$ & $1.00$ & $1.39$ & $0.48$ &
$3.70$ & $0.18\pm0.14\pm0.03$ \\
$\psi(4160)$ & $D\bar{D}/D^{\ast}\bar{D}^{\ast}$ & $0.08$ & $1.17$ &
$3.23$ & $0.27$ & $0.02\pm0.03\pm0.02$ \\
& $D\bar{D}^{\ast}/D^{\ast}\bar{D}^{\ast}$ & $0.16$ & $0.02$ & $1.40$ &
$0.03$ & $0.34\pm0.14\pm0.05$ \\
$X(4360)$ & $D\bar{D}/D^{\ast}\bar{D}^{\ast}$ & - & $0.40$ & $0.12$ &
$0.90$ & $0.14\pm0.12\pm0.03$ \\
& $D\bar{D}^{\ast}/D^{\ast}\bar{D}^{\ast}$ & - & $0.08$ & $0.64$ & $0.92$ &
$0.17\pm0.25\pm0.03$ \\
$\psi(4415)$ & $D\bar{D}/D^{\ast}\bar{D}^{\ast}$ & - & $1.54$ & $1.10$ &
$0.46$ &
$0.14\pm0.12\pm0.03$ \\
& $D\bar{D}^{\ast}/D^{\ast}\bar{D}^{\ast}$ & - & $0.28$ & $0.92$ & $0.18$ &
$0.17\pm0.25\pm0.03$ \\
\hline
\end{tabular}
\caption{\label{tab:ratios} Some ratios predicted theoretically by the
$^{3}P_{0}$ and the microscopic models. The comparison with the experimental
data is included.}
\end{center}
\end{table}

Therefore the full model has not solved the disagreement of the theoretical
calculation with the data and more theoretical and experimental work is needed
to solve the problem of the charmonium strong decay widths.

\acknowledgements{
This work has been partially funded by Ministerio de Ciencia y Tecnolog\'ia 
under Contract No. FPA2010-21750-C02-02, by the European Community-Research
Infrastructure Integrating Activity 'Study of Strongly Interacting Matter'
(HadronPhysics2 Grant No. 227431) and by the Spanish Ingenio-Consolider 2010
Program CPAN (CSD2007-00042).
}


%

}  


\end{document}